\documentclass[epj,referee]{svjour} 
\usepackage{amsfonts}
\usepackage{amssymb}
\usepackage{epsfig}
\begin{document}
\title{Effective temperature and jamming transition in dense, gently
sheared granular assemblies} 
\author{Fabricio Q. Potiguar\inst{1,2} \and Hern\'an A. Makse\inst{1}}
\institute{Levich Institute and Physics
Department, City College of New York, New
York, NY 10031 \and Universidade Federal do Cear\'a, Departamento de 
F\'\i sica, Campus do Pici, 60455-760, Fortaleza, Cear\'a, Brazil}
\abstract{
We present extensive computational results for the effective temperature, 
defined by the fluctuation-dissipation relation between the mean square 
displacement and the average displacement of grains, under the action of 
a weak, external perturbation, of a sheared, bi-disperse granular packing 
of compressible spheres. We study the dependence of this parameter on the 
shear rate and volume fractions, the type of particle and the observable 
in the fluctuation-dissipation relation. We find the same temperature 
for different tracer particles in the system. The temperature becomes 
independent on the shear rate for slow enough shear suggesting that it is 
the effective temperature of the jammed packing. However, we also show 
that the agreement of the effective temperature for different observables 
is only approximate, for very long times, suggesting that this defintion 
may not capture the full thermodynamics of the system. On the other hand, 
we find good agreement between the dynamical effective temperature and a 
compactivity calculated assuming that all jammed states are equiprobable. 
Therefore, this definition of temperature may capture an instance 
of the ergodic hypothesis for granular materials as proposed by theoretical 
formalisms for jamming. Finally, our simulations indicate 
that the average shear 
stress and apparent shear viscosity follow the usual relation with the shear 
rate for complex fluids. Our results show that the application of 
shear induces jamming in packings whose particles interact 
by tangential forces.
\PACS{
  {45.70.-n}{} \and
  {45.70.Vn}{}
 }
}

\authorrunning{Potiguar \and Makse}
\titlerunning{Effective temperature and jamming...}
\maketitle

\section{Introduction}
Granular materials are collections of macroscopic particles which
 interact, essentially, through inelastic collisions, repulsive
 contacts, and friction. This idea is simple and can be applied to a
 great variety of systems in nature, such as grains of sand, and
 assemblies of balls. Their importance is enormous to some industry
 segments, for example, the food industry \cite{Lar98}.

Even though granular materials may sound simple, their
physics is not. For reviews on general properties of such systems 
refer to the work by Jaeger and collaborators \cite{Beh96}.

The complex properties of granular systems make the development of a
general theory describing their behavior a difficult problem. A statistical 
mechanics approach was proposed by Edwards \cite{Edw89,mbe} 
to describe dense, slowly driven packings of rigid grains. The strong assumption
of Edwards is that the microscopic configurations, called {\em blocked} or 
{\em jammed}, that the system visits during its evolution are equally probable. 
By this hypothesis, Edwards derived a whole thermodynamics-like formalism for powders.

Makse and Kurchan \cite{Kur02} showed, for system of driven, compressible 
grains, that the quantity:
\begin{equation}
\label{compact}
X_E^{-1}=\frac{\partial S}{\partial E},
\end{equation}
where $S$ is the entropy of the powder and $E$, its energy, is the same, 
within error bars, as an effective temperature $T_{eff}$ obtained 
dynamically, as proposed by Cugliandolo et al. 
\cite{Cug97}. The quantity $X_E$ is the compactivity analogous to the 
compactivity of Edwards \cite{Edw89} where instead of using the energy $E$ 
it is used the volume $V$ of the powder. This effective
temperature, $T_{eff}$, is defined through the fluctuation-dissipation 
relation (FDR) between the mean square
displacement (MSD) and average displacement (AD) of grains in 
$z$-direction under the action of a weak external force $f$: 
\begin{equation}
\label{eff-temp}
\left<\left[z(t)-z(0)\right]^2\right>=2T_{eff}\frac{\left<z(t)-z(0)\right>}{f},
\end{equation}
the brackets indicate an average over time and particles. This parameter is also called
configurational temperature for it controls the structural relaxation of a jammed system.
This idea was studied in a variety of systems including glasses \cite{Cug97,Par97,Bar99,Sci01,Ber02-01}, 
models of powders \cite{Nic99} and foams \cite{Ono02}.

The purpose of the present paper is to deeper investigate this
result. We study the dependence of the configurational temperature
with the external parameters (volume
fraction and shear rate). We show that it does
not depend on the shear rate (for slow enough motion) and the
particular species of particle, a result that we believe characterizes 
the jamming, and provide initial evidence that the effective
temperature is a proper thermodynamic variable to describe the
system. We also discuss the dependence of $T_{eff}$ on
the observable used in the fluctuation-dissipation calculation.  It
has been recently shown that the effective temperature of dense
granular matter can be measured experimentally in a Couette cylinder
filled with spherical grains and slowly sheared \cite{swm}.  These
results together with the work performed by other groups provide
indications that the thermodynamics approach to dense granular matter
is possible.

We also analyze the role played by the granular temperature, as mesured by 
the average of kinetic energy. The results indicate that it is not a 
thermodynamical parameter for dense regimes due to its strong dependence on 
the external parameters. We also observe the behavior of
others macroscopic parameters of the packings, such as the average
shear stress and apparent shear viscosity. The first observable follows the 
usual Herschel-Bulkeley law. There seems to be a violation of this law for 
states in low densities 
where there is a frictional force between grains.  Besides, the denser
states develop a yield stress at low shear rates. These dense states are in 
the regime where the effective 
temperature is the relevant parameter. This implies the same correspondence 
already observed for glass-forming 
liquids \cite{Ber02-01}. The apparent shear viscosity displays the 
shear-thinning behavior of complex fluids, suggesting that powders may be 
treated as such fluids.

The organization of this work is as follows. In Section II we present
details of the procedures employed in the simulations.  In Section
III, the results for the diffusion coefficients, mobilities, defined by the MSD and AD, respectively,
 and effective temperatures are presented. Also in this section, we discuss 
the role of the granular temperature. The results of the average 
macroscopic quantities measured, shear stress and apparent shear viscosity, 
are shown in Section IV. Finally the conclusions are drawn in Section V.

\section{Simulation}

We use Molecular Dynamics (also called Distinct Element Method (DEM)
 \cite{Cun79}) to model our system. It is a  bi-disperse assembly of 1000 
spheres, 
half large, half small, with radii ratio of $R_S/R_L= 0.818$, which are 
initially randomly generated in a periodic cubic cell of size $L$.

The spheres interact via normal and tangential forces \cite{Johnson}. 
The first one is given by Hertz law, $F_n$. The second force
 is Mindlin's no-slip solution for the contact, $F_t$. The tangential force
 can increase up to the Coulomb threshold of friction $F_t=\mu F_n$ 
(where $\mu$ is the friction coefficient, typically $\mu=0.3$). If $F_t$ 
increases beyond $\mu F_n$, the grains slide and the contact is broken. We 
also simulated packings in which particles interact only through normal 
forces, and we refer to these cases no friction, or
 frictionless, systems. We notice that the frictionless case would model a 
system of compressed emulsions \cite{mbe,zha05}. Gravity is absent in all 
our simulations.

During the generation phase, no overlap between particles is allowed. 
We commonly have a initial volume fraction, $\phi$, of the
 order of $0.25$ in all our simulations.
Since our goal is to study dense systems, we should compress the
initial packing in order to attain a higher volume fraction. It is
known \cite{Mak00} that there is a jamming transition at a critical volume
fraction $\phi_c$, corresponding to the random close packing (RCP). Above
RCP there exists a mechanically stable solid-like state which is
characterized by a non-vanishing internal stress and coordination
number. Therefore, we choose three volume fractions, $\phi=0.6428$, 
$0.6623$, $0.7251$, above RCP to perform our studies.

We study the jamming transition by performing simulations also below RCP, 
i.e., in the fluid-like state. Such states can be prepared by expanding a
previously equilibrated system above RCP to decrease the volume
fraction up to the point where the pressure and the coordination
number vanish, as reported in \cite{Mak00}. We prepare three
fluid-like states with volume fractions given by $\phi=0.5999$, 
$0.6104$, $0.6318$.

\begin{figure}[h]
\centerline {\hbox{ \includegraphics*[width=.5\textwidth]{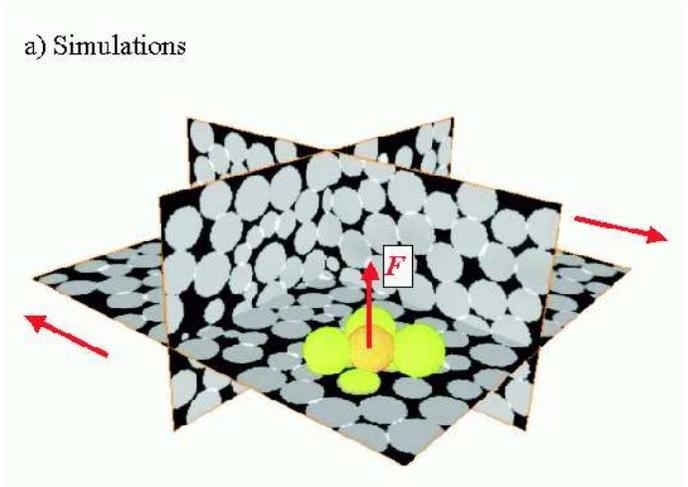}
} }
\caption{Detail of the simulations of grains of 100 microns
interacting via Hertz-Mindlin contact forces. A slow shear flow,
indicated by the horizontal arrows, is applied to the jammed system. We follow
the tracer particle trajectories to obtain the diffusivity. An
external force $F$ is then applied to the tracers in response to which
we measure the particle mobility. These dynamical measurements yield
an ``effective temperature'' obtained from an Einstein relation.}
\label{amira}
\end{figure}

The simulations involve the application of shear in the $x$-$y$ plane, 
at constant volume (see Fig. \ref{amira}), with $x$ the direction of the flow 
and $y$ the velocity gradient. We follow the trajectory particles in the
$z$-vorticity direction.  We use a modified version of the usual
Lees-Edwards boundary conditions \cite{Allen} which imposes a linear
velocity profile in the shear plane. Normal periodic boundary conditions are 
enforced in the
$z$-direction. Segregation was not observed in the time scale of our runs.
\vspace{2em}

For each volume fraction, 9 different shear rates were considered,
spanning a total of $5$ decades of magnitude:
$\dot\gamma=10^{-6},5\times10^{-6},10^{-5},5\times10^{-5},10^{-4},5\times10^{-4},
10^{-3},5\times10^{-3},10^{-2}$ s$^{-1}$. In our simulations time is given 
in seconds, length in meters, and mass in kilograms.

In what follows, we will calculate all the observables above and below
RCP and for frictional and frictionless particles
so that we will investigate the jamming transition at the RCP
limit. The results shown in the next sections will be only for large
particles, unless otherwise noticed. The results will be presented as functions of the shear rate,
 one curve for each volume fraction where applicable, in both frictional and frictionless cases.
 Moreover, since sheared systems enter a stationary state after an initial transient time \cite{Ber02-02,Yam98} 
and does not display aging, it is sufficient to
calculate all the following quantities as simple time averages. 
We calculate correlation and response functions with sets of overlapped data in the same run.
 We call this procedure window average. In practice it is the same as taking several time series
in different runs and averaging over them. It saves run time and storage space.

\section{Effective temperature}

We calculate the effective temperature, $T_{eff}$, of the granular
 system by measuring it through eq. (\ref{eff-temp}), as in
 \cite{Kur02}.  We investigate its dependence on the shear rate, and
 volume fraction. Next, we show that $T_{eff}$ does not depend on
 the particular species on which it is being measured.  Finally, we
 measure $T_{eff}$ by a different FDT commonly used in the glass
 literature, the self-intermediate scattering function (ISF) and averaging over
 the jammed configurations.
\vspace{2em}

\begin{figure}[h]
\rotatebox{0}{\epsfig{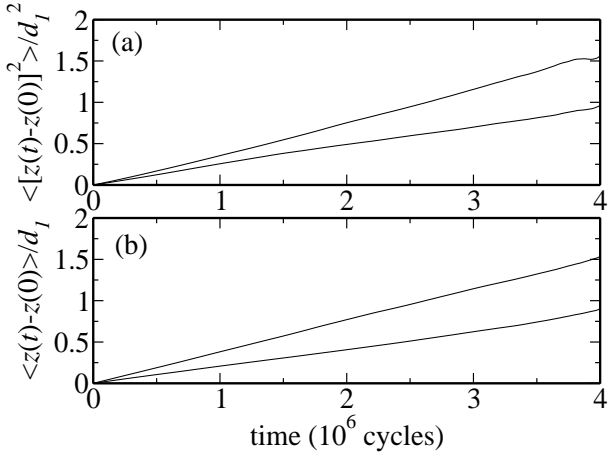}}
\caption{(a) MSD and (b) AD for large (lower) and small (upper) particles in units of large particle diameter, $d_1$.
}
\label{x2}
\end{figure}

\subsection{Diffusion, mobility and FDT}

The calculation of the diffusion coefficient in the $z$-direction,
 $D_z$, requires the evaluation of the MSD in this direction
 as a function of time:
\begin{equation}
\label{sqr-disp}
2D_zt=\left<\frac{1}{N}\sum_{i=1}^{N}\left[z_i(t)-z_i(0)\right]^2\right>,
\end{equation}
where $N$ is the number of particles considered, $z_i(t)$ is the
$z$-coordinate of the $i$-th particle at time $t$. 
The diffusion coefficient was calculated separately from the mobility
(see below). In all plots, lines connecting the points are only guides to
the eyes, unless otherwise noticed.  Figure \ref{x2}(a) shows the MSDs
 of the particles versus time from where the diffusion
constant is extracted.

\begin{figure}[h]
\includegraphics[width=8.0cm,height=6.0cm]{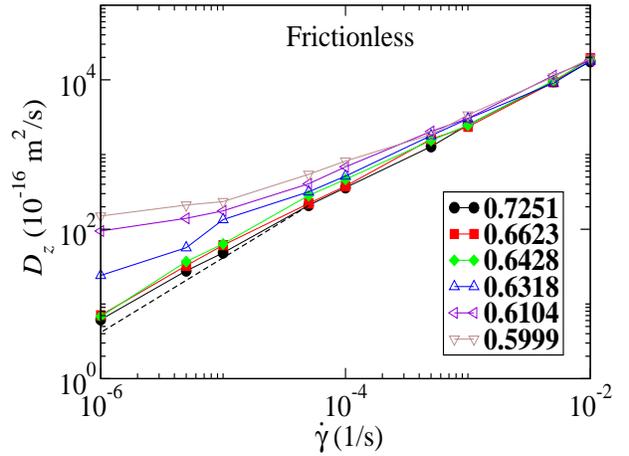}
\caption{Diffusion coefficient versus shear rate, no friction.}
\label{diff-nofr}
\end{figure}

\begin{figure}[h]
\rotatebox{0}{\epsfig{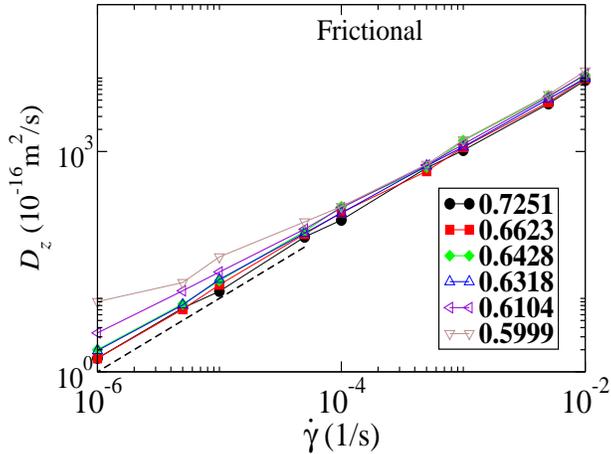}}
\caption{Diffusion coefficient versus shear rate, friction.}
\label{diff-fric}
\end{figure}

In Fig. \ref{diff-nofr} and \ref{diff-fric}, we show the results
for $D_z$ for frictionless and frictional cases, respectively. First,
 we notice that the high volume fraction curves go to zero in a similar fashion,
 suggesting that our system is jammed \cite{Dan01}. Also, as the volume fraction decreases,
particles have a higher diffusion for smaller shear rates. This is a
consequence of the larger space available for particles to move
 under the influence of enduring contacts. As we go up in shear
rate, this difference tends to disappear, the diffusion coefficient
becomes independent on the density, due to the larger effect
of inter-particle collisions.
\vspace{2em}

In the case of friction, particles still diffuse at lower densities
but, somewhat surprisingly, the diffusion coefficient becomes
independent on the density even for $\phi=0.6318$, a 
fluid-like state. This is a first indication that the presence of
friction between particles can induce the jamming transition for
systems that are below RCP. This is usually called a shear-induced 
jamming transition, see \cite{Far97,Bert02}.
\vspace{2em}

The mobility, $\chi_z$, is the response function
associated with the motion of particles under an applied external
force. To calculate $\chi_z$, a small constant force is applied
to each particle in order to induce a displacement in the 
$z$-direction. The mobility is taken as the slope of the curve of the
average $z$-displacement, normalized by the applied external
perturbation, versus time:
\begin{equation}
\label{mob}
\chi_{z}t=\frac{1}{f}\left<\frac{1}{N}\sum_{i=1}^{N}\epsilon_i\left[z_i(t)-z_i(0)\right]\right>,
\end{equation}
where $f$ is the applied force, and $\epsilon_i=\pm 1$ is a ``charge''
set to each particle $i$. This is a standard procedure employed
in numerical calculations to improve averages \cite{Hansen}. An
average over a distribution of these charges yields the desired
property. 
\vspace{2em}
\begin{figure}[h]
\rotatebox{0}{\epsfig{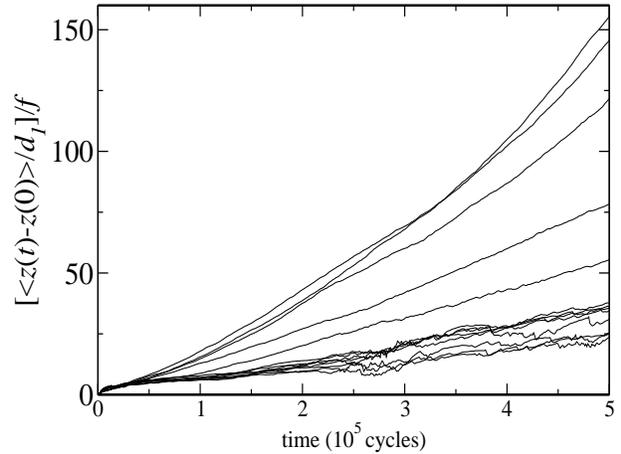}}
\caption{Average $z$-displacement (given in units of large
particle diameter) for force amplitudes of (from top
to bottom) $0.06$, $0.05$, $0.04$, $0.03$, $0.02$, $0.01$, $0.009$, $0.008$, 
$0.007$, $0.006$, $0.005$, $0.004$, $0.003$ N.}
\label{lin-reg-mob}
\end{figure}

Before calculating the mobility, we make
short runs in each state considered for, at least, 10 forces in order to
measure $\left<z(t)-z(0)\right>/f$.  The linear response regime is the
range of forces where all $z$-displacement curves, normalized by the
applied force, should overlap within numerical error. An example of this procedure
is shown in Fig. \ref{lin-reg-mob}. Here, we show several 
displacement curves, for different applied forces, in the case of
large, frictionless particles at volume fraction of $\phi=0.6428$ and
sheared at $\dot\gamma=5\times10^{-6}$ s$^{-1}$. We can see clearly
that for small forces, all $\left[z(t)-z(0)\right]/f$ curves (without
averaging) are within some well defined range, while at higher forces,
the curves starts to deviate from the linear regime. After this
regime is identified, we selected at least 5 values of the force to
perform the calculation in longer runs.
\vspace{2em}

\begin{figure}[h]
\rotatebox{0}{\epsfig{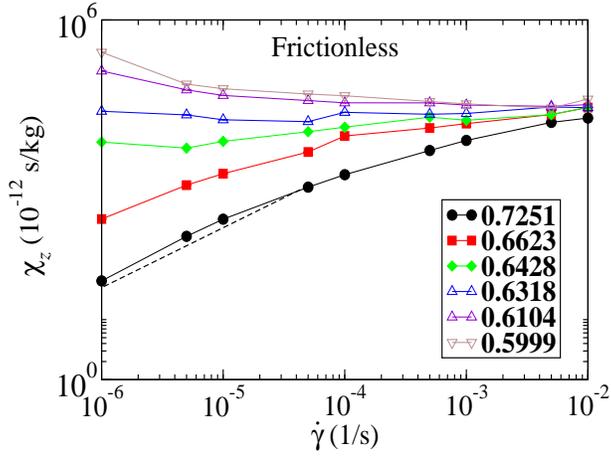}}
\caption{Mobility vs. shear rate, no friction.}
\label{moby-nofr}
\end{figure}

\begin{figure}[h]
\rotatebox{0}{\epsfig{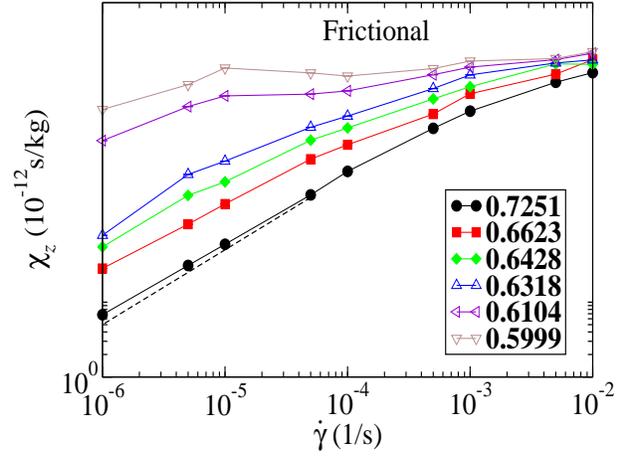}}
\caption{Mobility vs. shear rate, friction.}
\label{moby-fric}
\end{figure}

Figs. \ref{moby-nofr} and \ref{moby-fric} show the results for
$\chi_z$ for frictionless and frictional cases. Similarly to the case 
of diffusion, the mobility of the
grains increases with decreasing volume fraction. Also, the presence
of friction between grains diminishes the ability of particles to
move, hence yielding a smaller mobility for these packings.
The $\phi=0.6318$ friction packing has similar
dependence on $\dot\gamma$ as its denser counterparts, confirming the
suspected shear-induced jamming transition indicated in the
diffusivity plot of fig. \ref{diff-fric}.

A clear trend observed in the plots for $D_z$ and $\chi_z$ is that, at lower 
volume fractions and shear rate, both $D_z$ and $\chi_z$ have weaker 
dependence on shear rate, clearly 
observed in the no friction plots (and particularly, in the $\phi=0.6428$
and $\phi=0.6318$ cases). This fact indicates that these states are not 
jammed, since the ratio between $D_z$ and $\chi_z$, and consequently 
$T_{eff}$, would not be independent on $\dot\gamma$.
A similar trend is displayed in the friction plot, but only to the two
states at $\phi=0.5999$ and $\phi=0.6104$, and for shear rates above 
$\dot\gamma=10^{-5}$ s$^{-1}$. 

At higher volume fraction, we have a different picture. The dependence on the
shear rate is approximately the same for all curves in all plots. This already suggests the regime where 
$T_{eff}$ would be the relevant thermodynamical parameter.
In these figures we put a dashed line with unity slope in order to
indicate the expected behavior of both $D_z$ and $\chi_z$ with the
shear rate. We have close agreement in our results.
\vspace{2em}

In Fig. \ref{x2}a we plot the MSD in the $z$-direction while in Fig. \ref{x2}b,
 the AD in $z$ for large and small particles after the external force is applied.
 We find that the small tracer gives rise to a significantly larger diffusion
constant than the large tracer. The same behavior is obtained for the
mobility as can be seen in Fig. \ref{x2}b. This case was the $\phi=0.7251$, 
frictionless one sheared at $\dot\gamma=5\times10^{-6}$ s$^{-1}$. 
The perturbation was $f=0.09$. All lengths are in units of large
particle diameter, $d_1$. In fig. \ref{x2}a, we measured 
$D_z/d_1^2=2.295\times10^{-7}$ s$^{-1}$, large, and 
$D_z/d_1^2=3.995\times10^{-7}$ s$^{-1}$, small
particles. Figure \ref{x2}b gave 
$\chi_z/d_1=2.232\times10^{-6}$ s/(kg$\cdot$m), large, and
$\chi_z/d_1=3.741\times10^{-6}$ s/(kg$\cdot$m), small particles. 
 The diffusion constant and the mobility are inversely related to the grain 
sizes. For a Stokes fluid we would expect that the ratios of the mobilities and
diffusivities satisfy $D_s/D_l = d_l/d_s$, and $\chi_s/\chi_l =
d_l/d_s$ where $d_s$ and $d_l$ are the diameter of the small and large
particles respectively.

In our simulations, the ratio $d_l/d_s$ is equal to $1.22$. We have,
for example, the diffusion coefficients for the $\phi=0.7251$ friction case, 
$\dot\gamma=10^{-2}$ s$^{-1}$, are $D_l=9.317\times10^{-13}$ and
$D_s=1.149\times10^{-12}$, their ratio yields $1.233$, thus in good
agreement with the Stokes result. Moreover, the mobilities for the
state at $\phi=0.6104$ volume fraction, sheared at $\dot\gamma=5\times10^{-6}$
s$^{-1}$ without friction have the values $\chi_l=6.818\times10^{-8}$
and $\chi_s=8.664\times10^{-8}$, whose ratio gives $1.27$.  
\vspace{2em}

\begin{figure}[h]
\rotatebox{0}{\epsfig{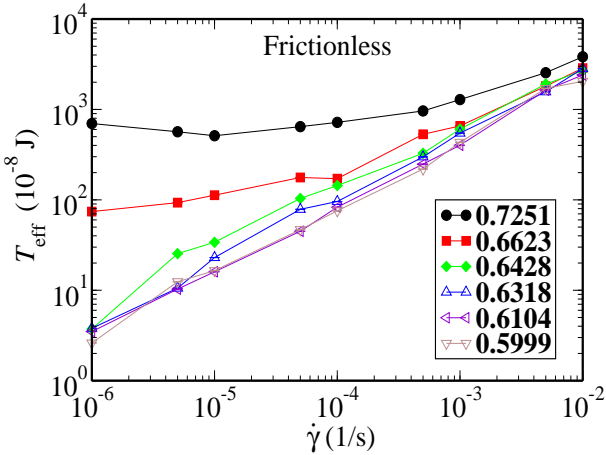}}
\caption{Effective temperature vs. shear rate, no friction.}
\label{teff-nofr}
\end{figure}

\begin{figure}[h]
\rotatebox{0}{\epsfig{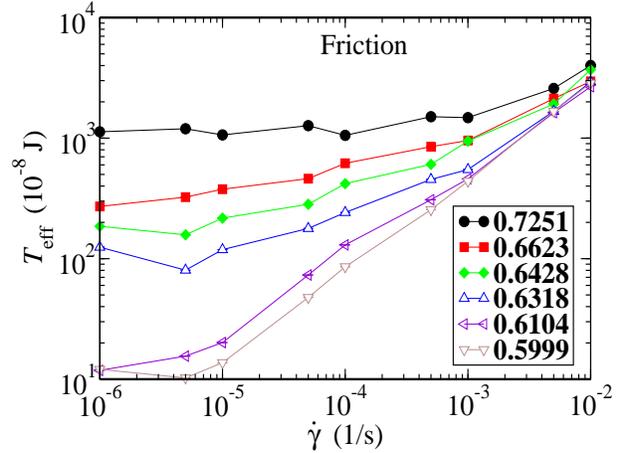}}
\caption{Effective temperature versus shear rate, friction.}
\label{teff-fric}
\end{figure}

Now we pass on to the results of effective temperature. They are given in
Figs. \ref{teff-nofr} and \ref{teff-fric}. We can see that at higher
volume fractions this effective temperature is approximately independent on the shear rate as
$\dot\gamma\rightarrow0$.  This confirms what was stated before about the range of jammed states.
 Also, it is an important result for it shows
that this is indeed an intrinsic property of the granular system,
since it does not depend on the driving strength. The jammed state can
be characterized by the independence of the effective temperature on
the shear rate since this behavior is not observed in the other cases
studied. 
\vspace{2em}

Two other conclusions may be drawn from these graphs. One is already expected
from the previous plots: the state $\phi=0.6318$ for the frictional case is
indeed jammed, although it is below RCP. This confirms that the application of shear 
can induce a jamming transition in a friction packing. The second conclusion 
is that the state $\phi=0.6428$
for frictionless particles is not jammed, even though it is above RCP. This
indicates that frictionless particles are unstable under shear, however small. 

\begin{figure}[h]
\rotatebox{0}{\epsfig{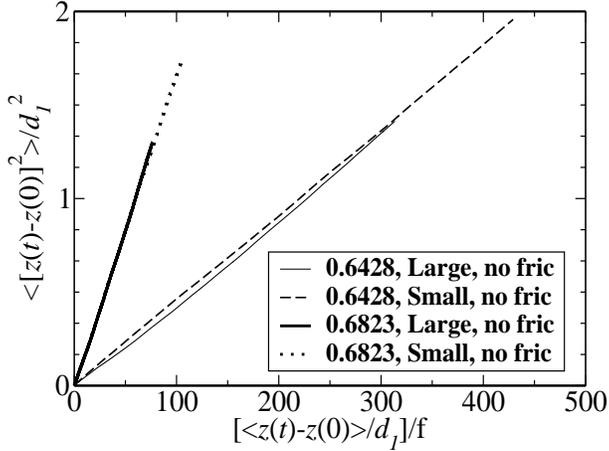}}
\caption{Parametric plot of the MSD by AD. The legend displays the
 cases studied. All lengths are given in units of large particle
diameters.}
\label{para-plot}
\end{figure}

Next, we test the zeroth law of thermodynamics for $T_{eff}$. In other words, 
if the effective temperature has a physical thermodynamic meaning, it should be the same for different
tracers in the medium. We test this by repeating the calculations of
the MSD and the AD for both types of particles in frictionless states 
$\phi=0.6428$ and $\phi=0.6623$. Figure \ref{para-plot} shows the parametric 
plot of the two quantities at the cases
considered, sheared at $\dot\gamma=5\times10^{-6}$ s$^{-1}$. The agreement is
clearly excellent. Although the curves for $\phi=0.6428$ are in close 
agreement, the results for $T_{eff}$ rules out the 
possibility that this state is jammed. This further suggests that $T_{eff}$ 
is a good thermodynamic variable for the system.
\vspace{2em}

\subsection{Self-Intermediate scattering function FDT}

Another important characteristic of the effective temperature is
that it should be independent of the particular observable (or
fluctuation-dissipation relation) used in its measurement. In other
words, we would like to know whether by measuring $T_{eff}$ using different
observables than those in (\ref{eff-temp}) we obtain the same result. This
independence on the observable was showed for supercooled liquids 
\cite{Ber02-01} and foams \cite{Ono02}.

Here we use a quantity which was extensively studied in the glass
literature, namely the self-intermediate scattering function (ISF) 
\cite{Bar99,Sci01,Ber02-01,Sas98,Kob95,Mia04}, defined as:
\begin{equation}
\label{inc-sca-fun}
C_{k_z}(t)=\left<\frac{1}{N}\sum_{i=1}^{N}
\exp\left\{i{k_z}\left[z_i(t)-z_i(0)\right]\right\}\right>
\end{equation}
where $k_z$ is the wave vector in the $z$-direction. This function is the
Fourier transform of the self part of the Van Hove correlation function at time $t$ \cite{Hansen}. 
We chose the wave vector $k_z$ as the peak of the static structure factor, which,
in our case, is $k_z=71.8 L^{-1}$.

In order to measure a FDR (\ref{inc-sca-fun}) we 
apply a small force in each particle $i$ of the form $F_i=-f k_z 
\epsilon_i \sin(k_zz_i)$, where $f$ is the amplitude of the force.
 After the perturbation is applied 
we measure the response function, defined as: 
\begin{equation}
\label{res-fun}
R_{k_z}(t)=\frac{1}{f}\left<\frac{1}{N}\sum_{i=1}^{N}\epsilon_i
\exp\left[ik_zz_i(t)\right]\right>.
\end{equation} 
The amplitude $f$ is the quantity which controls the intensity of the 
force. Therefore we set it to be small enough so that we measure 
$R_{k_z}$ in the linear response regime. As in the mobility 
calculation, we search for the linear response regime by measuring 
the response function for several forces before performing the full 
calculation. Also, we did not employ the window average in the 
calculation of (\ref{res-fun}), instead we averaged over different
 measurements, $20$, for each force.

The effective temperature is defined as the ratio between the real
parts of (\ref{inc-sca-fun}) and (\ref{res-fun}):
\begin{equation}
\label{eff-tem-02}
T_{eff}=-\frac{{\rm Re}\left[C_{k_z}(t)\right]}{{\rm
Re}\left[R_{k_z}(t)\right]}.
\end{equation}
\vspace{2em}

\begin{figure}[h]
\rotatebox{0}
{\epsfig{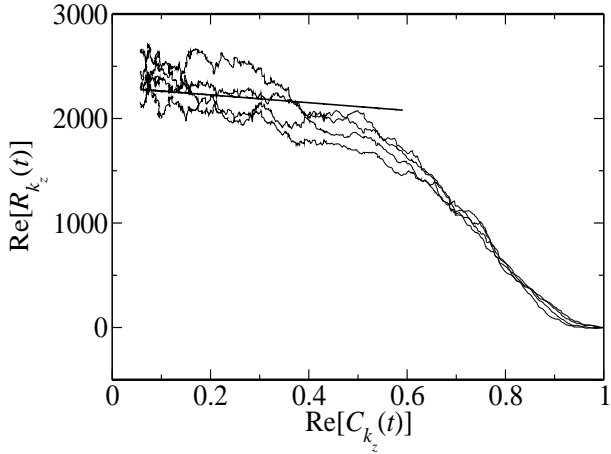}}
\caption{The effective temperature (straight line) calculated from
(\ref{eff-temp}) overlaps with fluctuation-dissipation plots from
(\ref{eff-tem-02}) in the very long time limit. The amplitudes of the 
applied perturbations are (from top to bottom) $5\times10^{-5}$, 
$6\times10^{-5}$, $7\times10^{-5}$, and $8\times10^{-5}$ N. 
This case is for frictionless particles, volume fraction of $\phi=0.6428$ 
and shear rate of $\dot\gamma=5\times10^{-6}$ s$^{-1}$.}
\label{teff-corr}
\end{figure}

We measured this effective temperature for the $\phi=0.6428$ frictionless state
 sheared at $\dot\gamma=5\times 10^{-6}$ s$^{-1}$. In order to test the 
equivalence of both temperatures, we plotted all results for 
(\ref{eff-tem-02}) parametrically and a straight line with slope 
$-{T_{eff}}^{-1}$, with its value given by the corresponding result from 
(\ref{eff-temp}). This is shown in Fig. \ref{teff-corr} for 
large particles. We see that the straight line has slope in the order of 
magnitude, only for very long times, as the one obtained from the ISF. Thus, 
it possibly indicates that $T_{eff}$ do not capture all the statistical 
mechanics of the jammed state.

\subsection{The role of kinetic energy}

The kinetic energy is defined as the sum of the velocity fluctuations of all 
particles:
\begin{equation}
\label{kin-ene}
2E_K=\sum_{i=1}^{N}m_i\left({{\bf v}_i-\left<{\bf
v}_i\right>}\right)^2,
\end{equation}
where $m_i$ is the $i$-th particle's mass and ${\bf v}_i$, is the velocity 
vector. 

As in thermal systems, we can define a parameter, called granular
kinetic temperature, $T_K$, which is the average of (\ref{kin-ene})
divided by $3N$.  Since the average velocity is zero in the $z$
direction, since there is no bias in this direction, we take $T_K$ as
the following expression:
\begin{equation}
\label{kin-temp}
T_K=\frac{1}{N}\sum_{i=1}^{N}m_i{v_z}^2.
\end{equation}
This is an estimate of the temperature of the fast modes of the system 
\cite{Cug97,Ber02-02} (since a granular system is athermal by
definition). In thermal systems, eq. (\ref{kin-temp}) yields the bath 
temperature. It is expected, in the dense limit, that $T_K$ does not share 
the same characteristics of $T_{eff}$. 

\begin{figure}[h]
\rotatebox{0}{\epsfig{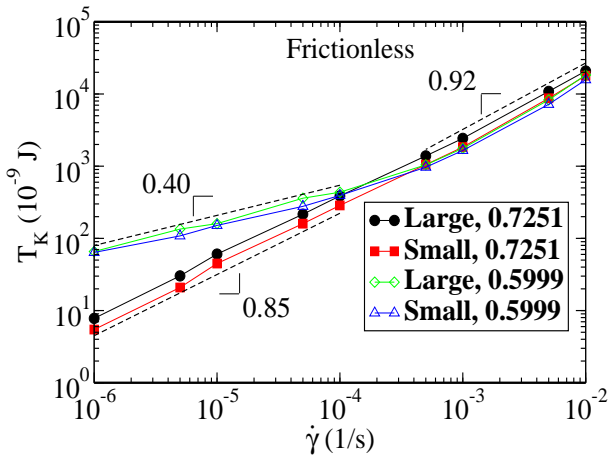}}
\caption{Average kinetic temperature vs. shear rate, no friction, for
large and small particles. The legend displays the cases showed.
 The dashed lines are fits to the curves and the numbers
are the slopes of these lines (explained in the text).}
\label{equi-nofr}
\end{figure}

In Fig. \ref{equi-nofr}, we show the kinetic temperature for both
 large and small particles for the highest and lowest volume fractions
 studied in the frictionless case. It is directly seen that such plots
 evidence the difference of $T_K$ for both species of particles. We also 
notice that the kinetic temperature has two different regimes. One at high 
shear rate, where it is almost density independent.  This is the fluid regime 
where the kinetic energy is dissipated at a smaller pace than in the dense
 case, this dissipation made mainly by collisions between grains. At low 
shear rates, the effect of volume fraction shows up. The denser the system, the
 smaller the kinetic temperature, as expected, since at high
 volume fractions there is a large number of contacts in the packing,
 which rapidly dissipate kinetic energy. In this case,
 the kinetic temperature depends on the shear rate again as a power
 law, but with a density dependent exponent.
\vspace{2em}

\begin{figure}[h]
\rotatebox{0}{\epsfig{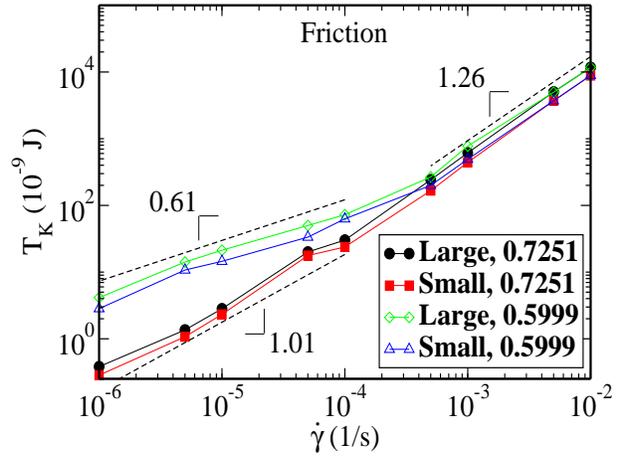}}
\caption{Average kinetic temperature versus shear rate, friction for large
and small particles. The legend indicates the cases showed.
 The dashed lines are fits to the curves and the numbers are the slopes of 
the lines (explained in the text).}
\label{equi-fric}
\end{figure}

Figure \ref{equi-fric} shows the kinetic temperature for large and small 
particles, in the lowest and
highest volume fractions considered here for frictional packings. The
same conclusions drawn from plot \ref{equi-nofr} can be taken here.
This parameter is indeed different for both particle species, confirming that
$T_K$ should not be used as a thermodynamical parameter for dense
systems. As in the previous graph, $T_K$ has two different regimes. One
for high shear rate, where it scale with $\dot\gamma$ as a power law
with exponent $1.26$. The second one, where the jammed states are,
this exponent is density dependent, the lowest being $0.61$ and the
highest $1.01$. 
\vspace{2em}

All dashed lines and their respectively slopes are fits
to the curve of $T_K$ for the total number of particles, without any
distinguishing from large or small. It should be noticed that the
results of \cite{Fei02} indicate that the kinetic temperature is
linearly dependent on the driving strength, the constant of
proportionality being a function of the density and elasticity of the
components of the mixture. Here we obtained a power law dependence
between the two quantities. It should be noticed, however, that the
previous results are for a granular bed, while ours is a 3D 
system. This difference may induce different dependencies in these
parameters.

A comparison between Figures \ref{equi-nofr} and \ref{equi-fric} with 
\ref{teff-nofr} and \ref{teff-fric}, respectively, implies that the effective 
temperature is always larger than the granular one in the jammed regime. 
We can also see that the behavior of $T_{eff}$ 
with volume fraction is different from the behavior of $T_K$ with 
$\phi$. The last one decreases with increasing $\phi$ while the former 
increases. Also, one can clearly see that the 
ratio of the values of $T_K$ for large and small particles is also 
constant, independent on the shear rate. This is the same observation 
that was made at the experiments of \cite{Fei02,Wil02}. Their result, 
which was obtained for a system in a fluid-like state, also applies to 
a dense one, at least to the component $T_{zz}$ of the temperature 
tensor (as defined in \cite{Biz99}). We did not studied the other 
components of this tensor.  Finally, we see that the kinetic 
temperature is always larger for large grains, a result also observed 
in \cite{Fei02,Wil02}.

The growth of the exponent of $T_K$ in $\dot\gamma$ with volume fraction can 
be seen in terms of the interaction 
between grains. Shearing a packing puts energy into it and the interactions 
dissipate this energy. As we lower the shear rate, enduring contacts build 
up as the main interaction mechanism. Since at lower volume fractions there 
are less contacts, diminishing $\dot\gamma$ does not cause a sharp 
drop on the average energy. For denser cases, the same reduction in shear 
rate causes the average energy to greatly lower its value due to the presence 
of more contacts. Therefore, $T_K$ drops faster with
 increasing $\phi$. Moreover,exponents are larger for friction packings due 
to the extra dissipation provided by tangential forces. 

\section{Calculation of the temperature via configurational averages}

So far we have shown the existance of an effective temperature for 
jammed granular materials and argue that this could be a thermodynamical
variable of the system because two different subsystems
characterized by particles of different sizes have the same value of this 
parameter, as shown in fig. \ref{para-plot}.
This is a kind of zero-th law for the system.
This property is not shared by, for instance, the
granular temperature, and therefore it could not be thought of as a 
physical variable describing the thermodynamics of grains.

We also showed that the effective temperature is within the range
of the temperature measured by other observables but only
for the very long time behavior of the system, as shown in 
Fig. \ref{teff-corr}.
Thus our results suggest that this temperature may not capture all the
thermodynamics of jamming.

Another test of the 
physical significance of $T_{eff}$ is to compare its value 
with a temperature obtained directly by an average over the 
jammed configurations.

If it was true that a thermodynamic framework could describe the
behavior of jammed systems, it stands to reason that the
compactivity $X_E$ of the granular packing given by eq. (\ref{compact})
should be measured from a 
dynamical experiment involving the exploration of the jammed states.
If the system is ergodic and amenable to a statistical
mechanical approach, then the temperature defined by the 
configurational average should be the same
as a dynamical measurement, as done for instance by the FDT.

To test this assumption is to test
the ergodic hypothesis for jammed matter.
Indeed, the statistical framework proposed by Edwards \cite{Edw89}
is based on the fact that the jammed 
configurations are all equiprobable.
This assumption is analogous to the basis under which
the micro-canonical ensemble of equilibrium statistical mechanics
is built. However, this assumption 
has been very controversial for granular matter since
grains are path dependent, frictional and non-conservative.
In the absence of a Hamiltonian to describe the dynamics,
there is no Liouville's theorem on which we can construct the 
statistical mechanics of grains from first principles.
Thus,the idea of ergodicity for 
granular materials has received much opposition.

The validity of this statement was analyzed through computer
simulations by Makse and Kurchan\cite{Kur02}. 
Our present results are an extension of these results.
Here we reproduce the discussion of this paper to make an explicit 
connection with previous work.
We will calculate the temperature from the entropy of the packing
with a mathematical construction which assumes the jammed
states to be equally probable.
The temperature calculated from the entropy of the packing is compared 
to the effective temperature obtained dynamically.
This is achieved via the procedure applied above
which has for an aim to probe each
static configuration by allowing the system to evolve at a very
slow shear rate. This is done
through the FDR which does not assume the flat average.

\subsection{Exploring the jammed configurations via a flat average.
Test of ergodicity: $T_{\mbox{\scriptsize eff}} = X_E$}
\label{ergodicity}

The independent method to study the configurational space of grains allows 
us to investigate the statistical properties of the jammed states 
available at a given energy and volume. In turn, it is investigated
 whether it is possible to relate the dynamical temperature obtained 
above via a diffusion-mobility protocol to the configurational compactivity
based on the flat average over jammed states.

In order to calculate $X_{\mbox{\scriptsize E}}$ and compare with
the obtained $T_{\mbox{\scriptsize eff}}$, it is needed to sample the
jammed configurations at a given energy and volume in an
{\it equiprobable} way. In order to do this, the jammed
configurations are sampled with the following probability distribution:
\begin{equation}
P_\nu \sim \exp[- E^\nu/T^* - E^\nu_{\mbox{\scriptsize
jammed}}/T_{\mbox{\scriptsize aux}}] \label{partition}
\end{equation}
Here the deformation energy $E$ corresponds to the  Hertzian
energy of deformation of the grains. 
The extra term added in Eq. (\ref{partition})
 allows the flat sampling of the jammed states to be perfomed. 
The jammed energy is such that it vanishes at the jammed configurations:
\begin{equation}
E_{\mbox{\scriptsize jammed}} \propto \sum_i
\left|\vec{F}_i\right|^2,
\end{equation}
where $\vec{F}_i$ is the total force exerted on a particle by its neighbors.
 The main property of this energy is that it should be zero at jamming.

\begin{figure}
\centerline {\hbox{
\includegraphics*[width=.5\textwidth]{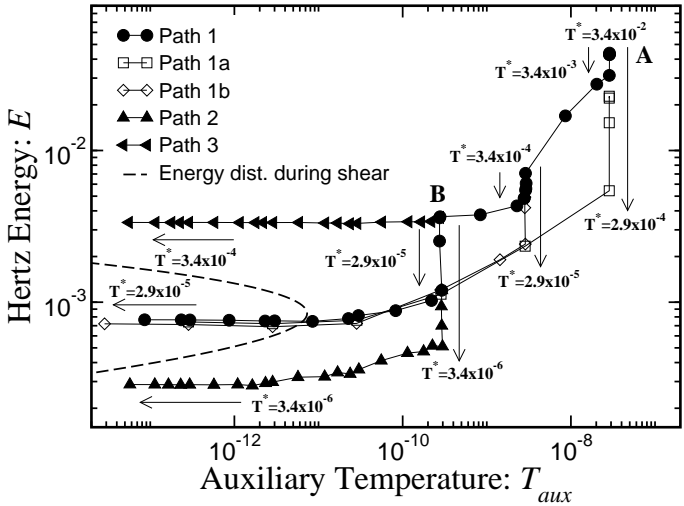} } }
\caption{ Annealing procedure to calculate  $X_E$ at different
elastic compressional energies.  We plot the elastic energy vs
$T_{\mbox{\scriptsize aux}}$ during  annealing together with
the distribution of elastic energies obtained during shear (dashed
curve, mean value $\langle E \rangle = 8.4\times 10^{-4}$). 
At the end of the annealing we find that the energy is
very similar to the average energy for the sheared system
only when we set 
$T^* = X_E$.
} \label{annealing}
\end{figure}

Two ``bath'' temperatures are introduced (these temperatures are
$\sim 10^{14}$ times the room temperature) which allow the
exploration of the configuration space and the calculation of the entropy of
 the packing assuming a flat average over the jammed configurations. 
Equilibrium MD simulations are employed with two
auxiliary ``bath'' temperatures $(T^*,T_{\mbox{\scriptsize
aux}})$, corresponding to the distribution
(\ref{partition}). Annealing $T_{\mbox{\scriptsize aux}}$ to zero
selects the jammed configurations ($E_{\mbox{\scriptsize
jammed}}=0$), while $T^*$ fixes the energy $E$.

In practice, the equilibrium MD simulations are carried out with a modified
potential energy:

\begin{equation}
U = \frac{T_{\mbox{\scriptsize aux}}}{T^*} E +
E_{\mbox{\scriptsize jammed}},
\label{potential}
\end{equation}
and calculate the force on each particle from $\vec{F} = -
\vec{\nabla} U$. 

The auxiliary temperature
$T_{\mbox{\scriptsize aux}}$ is controlled by a thermostat which
adjusts the velocities of the particles to a kinetic energy
determined by $T_{\mbox{\scriptsize aux}}$. We start by
equilibrating the system at high temperatures
($T_{\mbox{\scriptsize aux}}$ and $T^*$ $\sim \infty$) and anneal
slowly the value $T_{\mbox{\scriptsize aux}}$ to zero and tune
$T^*$ so as to reach the  value of $E$ that corresponds to the
average deformation energy obtained during shear.

The partition function is
\begin{equation}
Z = \sum_\nu \exp[- E^\nu/T^* - E^{\nu}_{\mbox{\scriptsize
jammed}}/T_{\mbox{\scriptsize aux}}], \label{partition2}
\end{equation}
from where the compactivity reads:
\begin{equation}
T^* = \frac{\partial E}{\partial S} \stackrel
{\mbox{\scriptsize $T_{\mbox{\scriptsize aux}}\to 0$}}{\longrightarrow} X_E,
\end{equation}
Thus at the end of the annealing process ($T_{\mbox{\scriptsize
aux}}\to 0$), $T^*(E)=X_{E}(E)$, since in this limit the 
sampled configurations have a vanishing number of moving 
particles at a given $E$.

Since a calculation of the force from a
potential energy, eq. (\ref{potential}), is required, 
only conservative systems can be studied with the method. 
Therefore it is not possible to test these
ideas in a granular system with friction for it is a path-dependent system.
Thus we focus our calculations on a system of frictionless particles. 

A system of {\it frictionless} viscoelastic spherical particles could be
thought of as a model of compressed emulsions
\cite{lacasse,durian}, see also \cite{faraday} for details. Even
though they can be modeled in this way, an important difference
arises in the inter-droplet forces, which are not given in terms of
the bulk elasticity, as they are in the Hertz theory.
Instead, forces are given by the principles of interfacial
mechanics \cite{princen}. For small deformations with respect to
the droplet surface area, the energy of the applied stress is
presumed to be stored in the deformation of the surface. The
simplest approximation considers an energy of deformation which is
quadratic in the area of deformation \cite{princen}, analogous to
a harmonic oscillator potential which describes a spring
satisfying Hooke's law.
More elaborated models have been proposed, and in a recent
study the force law was calculated experimentally 
using confocal microscopic images of emulsions \cite{zha05}.
Here, we will avoid issues of path-dependency
introduced by the interparticle
tangential forces, by excluding these forces 
from the calculation ($F_t = 0$). Because there
are no transverse forces, the grains slip without resistance and
this procedure essentially mimics the path to jammed states for
the compressed emulsion system.

Thus, the MD model of granular materials is adapted to describe the
system of compressed emulsions by only excluding the transversal
forces (tangential elasticity and Coulomb friction). The
continuous liquid phase is modeled in its simplest form, as a
viscous drag force acting on every droplet, proportional to its velocity.
The systems under investigation have exponentially large (in the
number of particles) number of stable states jammed at zero bath
temperature. Thus, a meaningful annealing procedure is
only possible for very small system sizes with the present computational power.
Therefore, the system of study has 200 particles only.

The calculations for the effective temperature for this system 
have been done in \cite{Kur02}. The effective temperature is 
$T_{eff}=2.8\times10^{-5}$, the mean pressure $P=10$ MPa,
the volume fraction $\phi=0.66$ and the mean compressional energy
$\left<E\right>=8.4\times10^{-4}$ J. All these values have been obtained 
during the shear of the material.

In the calculation of the compactivity
for the same system of 200 frictionless particles at the same pressure 
and volume fraction, there is no dynamics
so the viscous forces of dissipation are completely disregarded.
At the end of the annealing protocol the compactivity at a given deformation
energy can be obtained as illustrated in Fig. \ref{annealing}.

In the begining, the system is equilibrated for $40\times 10^6$ iterations at
high auxiliary temperatures, such as at 
point A in Fig. \ref{annealing}:
$(T^*=3.4\times 10^{-2}, T_{\mbox{\scriptsize aux}}=3\times
10^{-8})$. Different annealing paths are taken for $T^*$, one decreases its 
value to $T^*=2.9\times10^{-4}$ (open squares, path 1a) and the other anneals 
it to $T^*=3.4\times10^{-3}$ (filled circles, path 1). Path 1a is followed 
until the value of $T^*$ reaches $2.9\times10^{-5}$. Path 1 continues until 
both temperatures are annealed to the point where this path is split again: 
one branch anneals $T^*$ to $2.9\times10^{-5}$ (open diamonds, path 1b), 
joining the line of path 1a, while the other anneals to 
$T^*=3.4\times10^{-4}$. From this point on, $T^*$ is kept constant in paths 1a
 and 1b, while annealing $T_{\mbox{\scriptsize aux}}$ to zero. Path 1 
continues until it reaches 
point B, with $T_{\mbox{\scriptsize aux}}=3\times10^{-10})$, 
where it is split for the last time in three different branches: 
one keeps $T^*$ constant at $3.4\times10^{-4}$ (filled right triangles, path 3)
, the second anneals $T^*$ to the value $2.9\times10^{-5}$, the last one
 anneals $T^*$ to $3.4\times10^{-6}$ (filled triangles, path 2). 
When $T^* = T_{\mbox{\scriptsize eff}}$ (Paths
1, 1a and 1b), the final elastic compressional energy value when $
T_{\mbox{\scriptsize aux}}\to 0$ falls inside the distribution of
energies obtained, and it is very close to the mean value of the
elastic energy during shear $\langle E \rangle$. 
The remarkably result is that this implies that the compactivity and the 
effective temperature obtained dynamically are found to coincide
to within computational error,
\begin{equation}
X_E \approx T_{\mbox{\scriptsize eff}}.
\end{equation}

For other values of $T^* \ne T_{\mbox{\scriptsize eff}}$ the final $E$ 
falls out of the distribution obtained during shear (Paths 2 and 3).

This provides evidence for the validity of the effective temperature as a 
dynamical estimate of the compactivity, and more importantly, justifies the
use of the statistical measurements presented in
characterizing the macroscopic properties of the jammed system.
Since $T_{eff}$ does not assume the flatness of the
jammed states because it is obtained dynamically,
these results indicate that, at least in the present system, the ergodic
hypothesis for granular matter is a plausible approximation to the
dynamics of the system.

The conclusion is that the jammed configurations
explored during shear are sampled in an equiprobable way as
required by the ergodic principle. Moreover, the dynamical measurement of
compactivity renders the thermodynamic approach amenable to
experimental investigations \cite{swm}.

It should be noted though that this is only
an approximation, and it is valid only when the granular material
is pretreated so that it is in a reversible state.
In computer simulations, the reversible state is achieved by
simulating packings without friction as explained before.
In the case of friction, it is still possible 
to generate packings which are reversible and amenable to a
statistical description.
This problem has been treated before in \cite{zha05,Mak04}.
Experimentally, it has been also shown 
that granular matter should show reversibility if it is 
properly prepared, even when system has friction and path-dependent forces.
The seminal experiments of the Chicago group of 
column tapping \cite{now97,now98} clearly indicate the existance of these
reversible jammed states in granular materials. Moreover
these experiments have been reproduced by other
groups as well \cite{phi02} and under other conditions such
as oscillatory pressure \cite{bru05}.
These reversible states are the ones for which the 
statistical mechanics is possible.
Notice that this assumption preclude other simple systems such as
a pile of grains. A sandpile, for instance, is not in a reversible
state since any perturbation will deform the pile.
Moreover, there is a whole branch of irreversible states
that have been obtained in the Chicago experiments and in the experiments of 
Brujic et al. \cite{bru05}, before the reversibility is achieved.

The present thermodynamics is not applied to such systems.
Other theories, such as the fragile matter theory of Cates et al. \cite{Cat98}
might be able to capture this irreversible states of granular matter.

\section{System properties}
In this section, we investigate how other properties of our system
behave as a function of the shear rate and density. The properties
observed were the average potential energy, pressure, and shear stress during shear 
(only shear stress from now on) and apparent shear viscosity. We want to know how 
jamming affects these properties. We only show results for the shear stress and the
 viscosity, since the potential energy and pressure behave qualitatively as the shear stress.
All the properties are calculated when the system enters 
the steady state. Therefore, they are given as simple time averages.

\subsection{Shear stress}

The shear stress is defined as the $xy$-component of the stress tensor
$\sigma_{ij}$. In fact, we measured the shear stress which arises from
compression and do not include the contribution from momentum exchange
between grains.  In complex fluids, the average shear stress is
usually fitted to a power law, the Herschel-Bulkeley equation \cite{Hol93}:
\begin{equation}
\label{power-law-str}
\left<\sigma_{xy}\right>=\sigma_0+A\dot\gamma^n,
\end{equation}
where $\sigma_0$ is the yield stress and $A$ is a constant 
\cite{Ber02-02,Sol97,Sol98}.
\vspace{2em}

\begin{figure}[h]
\rotatebox{0}{\epsfig{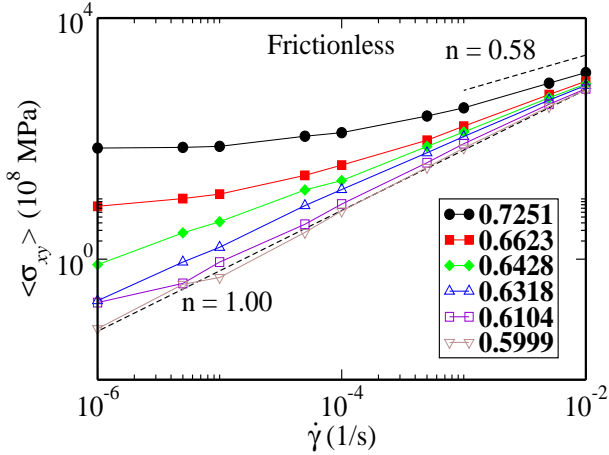}}
\caption{Average shear stress vs. shear rate, no friction. The plateaus are 
the evidence of a jamming transition. The dashed lines are explained in the 
text. The numbers are their slopes.}
\label{shst-nofr}
\end{figure}

In Fig. \ref{shst-nofr} is shown the results for $\sigma_{ij}$ for the 
frictionless case. It is observed a plateau at higher volume fractions 
and small shear rates. The plateaus correspond to the yield stress 
$\sigma_0$ in (\ref{power-law-str}). It is not surprising to observe 
a non-vanishing yield stress since equilibration of such packings implies 
that. What is interesting is the absence of an yield stress for the
 $\phi=0.6428$ state, initially in the solid-like state. This confirms that 
this system is unstable under shear and it is not truly jammed. Also, we 
let the system relax after shearing and observe that, indeed, 
$\left<\sigma_{xy}\right>\rightarrow0$. The existence of a yield stress can 
be considered 
as an instance of the jamming transition \cite{Ber02-02,Sol97,Sol98,Mat01,Ohe03}. 
\vspace{2em}

\begin{figure}[h]
\rotatebox{0}{\epsfig{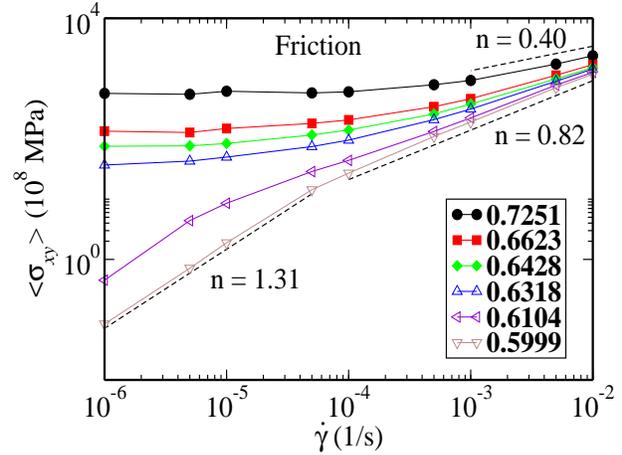}}
\caption{Average shear stress versus shear sate, friction. As in fig. 
\ref{shst-nofr} the plateaus correspond to jammed states. Dashed lines, 
showed with the correspondent slopes, are explained in the text.}
\label{shst-fric}
\end{figure}

As we go up in shear rate, the stress increases as a power law whose
exponent depends on the volume fraction (in the cases of the thermal
systems, this exponent depends on the bath temperature 
\cite{Sol97,Sol98}). The upper dashed line in \ref{shst-nofr} 
is a fit to the last $3$ points in the flow curve for the $\phi=0.7251$ case. 
The lower curves do not display a yield 
stress, but they match the opposite limit of the phenomenology of the 
shear stress, the power-law-fluid regime 
\cite{Ber02-02,Sol97,Sol98}. The lower dashed line is a fit to the flow 
curve of the $\phi=0.5999$ state, showing that this state corresponds to a 
Newtonian fluid. The two middle curves match the behavior of power-law fluid, 
with the exponent $n$ of (\ref{power-law-str}) density dependent.

In fig. \ref{shst-fric} we have the frictional results for 
$\left<\sigma_{xy}\right>$. We notice that there are four states that 
develop an yield stress, the same four that display a plateau in the
 effective temperature results.  As in the frictionless case, at high 
shear rate the main contribution to the shear stress is the second 
term on right hand side of (\ref{power-law-str}). Again the exponent 
$n$ depends on volume fraction. Its lowest value comes from the 
slope of the $3$ last points in the $\left<\sigma_{xy}\right>$ curve for the 
$\phi=0.7251$. On the other hand, it seems that, at lower 
volume fractions, the systems displays two different regimes, without 
an yield stress, in conflict with what is expected from the 
Herschel-Bulkeley equation. These two regimes are 
exemplified by the two lower dashed lines which are fits to 
the first $4$ points and to the last 5 points in the 
$\phi=0.5999$ flow curve. We would expect this curve to have a unity slope 
and this to be the Newtonian regime 
This fact deserves more investigation, 
since friction is inherent to grains, and its presence could alter 
significantly the behavior of the shear stress at low shear rate. The 
important conclusion is that at high shear rate the exponent $n$ is 
still density dependent and increases with smaller density, eventually 
reaching $1$.  Moreover, we also expect that at higher shear rates, 
the average shear stress should scale as 
$\left<\sigma_{xy}\right>\sim\dot\gamma^2$, the well-known Bagnold 
scaling \cite{Bag54}. 

By looking at the results for $T_{eff}$, the plots for shear stress 
confirm the two conclusion drawn before: instability of frictionless 
systems at densities up to $\phi=0.64$, at least (this is evidence of the 
fragility concept proposed to describe granular matter by Cates et 
al. \cite{Cat98}) and the shear-induced jamming transition. We can see 
that the plateaus in the shear stress (the yield stress), in the 
frictionless and frictional cases, correspond to the plateaus in the 
plots for $T_{eff}$.  Therefore, these states are the jammed states of 
our system.  In this way, we can say that the appearance of a 
well-defined effective temperature is a consequence of the development 
of an yield stress \cite{Ber02-01,Ber02-02,Mat01,Ohe03}. 

The packing instability can be explained by the fact that when shear is 
stopped, the grains become less compressed by pushing one another in 
order to attain a minimum of potential energy. While they push each 
other, they also slide related to their neighbors, breaking the 
contacts. When all, or most, of these contacts are broken all quantities that 
depend on the overlap between grains (potential energy, pressure, 
shear stress), should drop to zero, since the amount of compression is 
negligible. This relaxation does not happen at higher $\phi$ because 
there is no available space for the grains to fully relax. 

In the friction packings, the same mechanism of relaxation is 
present. This difference in the flow curves is a consequence of the fact
 that the remaining normal 
forces, responsible for separating the particles, are not strong 
enough to break the contacts, which are supported by the tangential 
forces. Therefore, we say that our system undergoes a shear-induced 
jamming transition in the case $\phi=0.6318$. The shear induces 
contacts between particles and these shear-induced contacts survive 
after the shear is stopped, providing the system an yield stress. 

\subsection{Apparent shear viscosity}
The last of the macroscopic observables probed is the apparent shear
viscosity of the system, $\eta$, defined as:
\begin{equation}
\label{visc}
\eta=\frac{\left<\sigma_{xy}\right>}{\dot\gamma}.
\end{equation}
Using eq. (\ref{power-law-str}), we can rewrite this expression as 
\cite{Ber02-02}:
\begin{equation}
\label{shr-thng}
\eta=A\dot\gamma^{-\alpha(\phi)}.
\end{equation}
From this equation we can see that, according to the value of $n$, the 
system may display three different regimes: shear-thinning 
($\alpha(\phi)>0$), Newtonian ($\alpha(\phi)=0$) or shear-thickening 
($\alpha(\phi)<0$).  Measurements of viscosity have been performed in 
a variety of complex fluids, for example glasses 
\cite{Yam98,Ber02-02}, foams \cite{Sol97,Sol98}, and colloids \cite{Tra01,Bert02,Lee03,Bonn02}. 
In some of these systems, a shear-thinning 
behavior, with exponents between $0.6$ and $1.0$, was observed 
\cite{Ber02-02,Yam98,Bonn02}, and in others both shear-thinning and 
shear-thickening are present \cite{Tra01,Lee03}. 
\vspace{2em}

\begin{figure}[h]
\rotatebox{0}{\epsfig{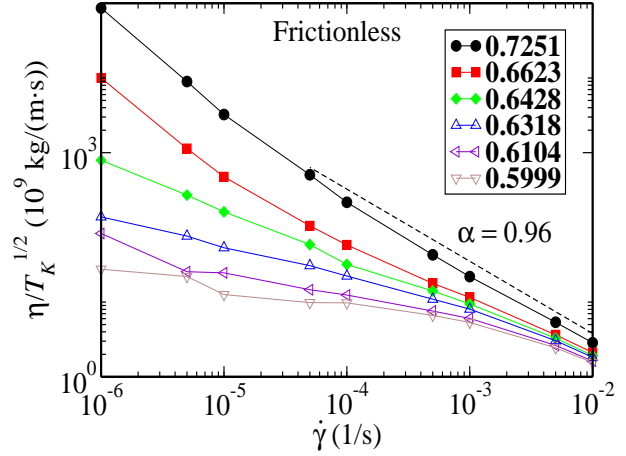}}
\caption{$\eta$ normalized by ${T_K}^{1/2}$, frictionless case. 
The slope of the dashed line indicate the regime we can find the system.}
\label{visc-temp-nofric}
\end{figure}

The calculation of (\ref{visc}) is not as direct as it seems. First of 
all it should be noticed that in the fluid state 
the viscosity scales as $T_K^{1/2}$ \cite{Biz99,Boc01}.  The reason 
behind this dependence is that in these cases, the grain movement is a 
consequence of the application of shear, which inputs energy in the 
system. This energy input, for instance, increases the kinetic energy 
of the grains. Therefore, the shear rate and the kinetic temperature 
(\ref{kin-temp}) are coupled. Since this input in energy leads to higher 
$T_K$, more collision take place and, consequently, the 
system reaches a higher stress. We should, then, plot the viscosity 
normalized by $T_K^{1/2}$ in order to disregard this effect on $\eta$. 
\vspace{2em}

\begin{figure}[h]
\rotatebox{0}{\epsfig{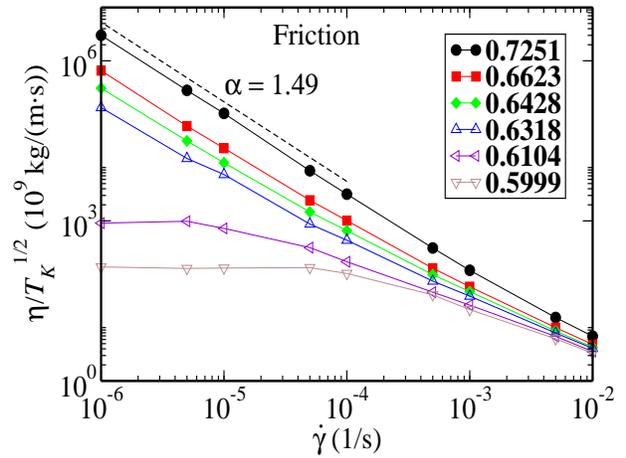}}
\caption{$\eta$ normalized by ${T_K}^{1/2}$, friction case. 
The Newtonian regime is clearer here. The dashed line is explained in the 
text.}
\label{visc-temp-fric}
\end{figure}

In fig. \ref{visc-temp-nofric} we plot $\eta/T_K^{1/2}$ for frictionless 
packings. We see that at low shear rate and low volume 
fractions we have approximately Newtonian behavior, already expected by 
the result from the shear stress, fig. \ref{shst-nofr}. This is the effect 
of the high velocities imposed on the system at this regime. Particles move so 
fast that they do not seem to ``see'' each other and then the whole 
viscosity drops. At higher volume fractions we clearly observe a 
shear-thinning behavior. We show a fit to the last $5$ points for the 
$\phi=0.7251$ curve. The slope is inside the range reported for many soft 
materials \cite{Ber02-02,Yam98,Bonn02}.

These regimes are more clearly observed in the friction case, 
fig. \ref{visc-temp-fric}. At lower densities and smaller shear rates, 
we have a Newtonian behavior with $\alpha$ close to zero. The 
situation is a bit different at higher densities and low shear 
rates. The slope of the fit (dashed 
line) for the first $5$ points in the $\phi=0.7251$ curve is outside the 
known range for complex fluids. Again, friction may be playing a major 
 role and more investigation is needed to elucidate this matter. 
It is important to check the behavior in 
the fluid regime, states with $\phi=0.5999$, $0.6104$ and $0.6318$. 
In the first two of them, we have a Newtonian behavior as expected for 
$\eta/T_K^{1/2}$. The third one displays shear-thinning since it is jammed, 
see figs. \ref{shst-nofr} and  \ref{shst-fric}. The results for the other 
states should be taken carefully, especially 
regarding the values of the exponents $\alpha$. But the trend is 
correct, the system should display shear-thinning at high shear rate 
as a result of jamming.

\section{Conclusions}
In this paper we study the effective temperature of sheared,
bi-disperse, dense granular system in three dimensions as a function
of the shear rate and volume fraction. We consider systems in which grains 
interact by purely repulsive forces, frictionless case, and others that they 
interact by normal and tangential forces, friction case.

The effective temperature (\ref{eff-temp}) is shown to be independent on the 
shear rate at volume fractions of
the order, and above, of $\phi=0.68$ for the frictionless case, and $\phi=0.63$
for the friction case. This effective temperature is shown to be equal
to both large and small particles, indicating that such parameter is
at least consistent with a zero-th law of thermodynamics for grains.
It is shown that $T_{eff}$ has some dependence on the observable as we had only 
approximate agreement with a temperature measured from the ISF.
 However, we present enough evidence pointing toward the 
fact that the effective temperature is an intrinsic property of the system 
and, along with experimental results \cite{swm}, that a thermodynamics of 
granular materials is feasible.

The fact that slow relaxation modes can be
characterized by a temperature raises the question of the
existence of a form of ergodicity for the structural motion,
allowing a construction of a statistical mechanics ensemble for
the slow motion of the grains. This argument leads us back to the
ideas of the thermodynamics of jammed states. In parallel to these
dynamical measurements, the same information is drawn from the
system by a flat statistical average over the jammed
configurations. Once all the static configurations have been
visited by the system, the compactivity $X_E$ can be calculated
from the statistics of the canonical ensemble of the jammed
states. The logarithm of the available configurations at a given
energy and volume reveals the entropy, from which the compactivity
is calculated. Our explicit computation shows that the 
temperature arising from the Einstein relation (\ref{eff-temp}) can be 
understood in terms of the configurational compactivity $X_{E}$ arising from
the statistical ensemble of jammed states. 

We also investigate the role played by the kinetic energy in the system.
The granular temperature does not share the properties of $T_{eff}$. It has a 
power-law dependence on the shear rate with exponents that increase with 
volume fraction. This is a direct consequence of the enduring contacts 
between grains.

The jamming transition is observed in the behavior of other observable
we studied, namely shear stress and apparent shear
viscosity. The states with volume
fractions higher than $\phi=0.68$ and $\phi=0.63$ in the frictionless and
friction cases develop a yield stress.  This is clear indication that the 
system jammed. Moreover, we showed that our system obeys the
Herschel-Bulkeley law (\ref{power-law-str}). We observed that the exponent 
is density dependent (in analogy with other soft materials, where it is
proportional to the external bath temperature). This parameter tends,
at high shear rate and low volume fraction, to $1$, in the
frictionless case, where the system behaves like a power-law fluid
 \cite{Ber02-02}. We did not observe this in the friction case. In fact
at lowest volume fraction, the system displays two different
exponents, and no yield stress. We attribute this to the possibility that 
friction between particle could invalidate (\ref{power-law-str}) for grains 
in some range of $\phi$.

The apparent viscosity, normalized by the square root of the granular
 temperature, displays the predicted behavior of shear-thinning observed for
others complex fluids at high volume fractions
 \cite{Ber02-02,Yam98,Biz99,Bonn02}. It depends as a power law on
$\dot\gamma$ with an exponent which depends on the volume fraction,
and approaches $-1.0$ as $\phi$ increases in the frictionless
case. This exponent is characteristic of complex fluids in jammed
states. The system shows a Newtonian behavior at low volume fraction
and shear rate. In the friction case, the Newtonian behavior is even
clearer, but we measured an exponent in eq. (\ref{shr-thng}) 
which is outside the known range for complex fluids, and
this may be, again, the presence of friction. 
However, the trends we obtained are important. We
observe a shear-thinning behavior for the system as long as it stays
jammed. All these results led to the
conclusion that by applying shear to a granular packing we can induce
a jamming transition (observed for the friction case at volume
fraction of $\phi=0.6318$) or we can unjam a system (observed for the
frictionless case at volume fraction of $\phi=0.6428$).

\section*{Acknowledgments}
We thank Mark Shattuck for many valuable discussions.  This work is
supported by the National Science Foundation, the Department of Energy
and CAPES (Brazilian agency).

\end{document}